\documentclass[]{interact}

\usepackage[utf8]{inputenc}
\usepackage[T1]{fontenc}
\usepackage[english]{babel}
\usepackage{textcomp}
\usepackage{csquotes}
\usepackage{latexsym, amsmath, amssymb}
\usepackage{booktabs, multirow}

\usepackage{graphicx}
\usepackage{dcolumn}

\usepackage[numbers,sort&compress]{natbib}
\bibpunct[, ]{[}{]}{,}{n}{,}{,}

\usepackage{color}

\usepackage[caption=false]{subfig}

\usepackage{mathtools}
\usepackage{exscale, relsize}
\usepackage[retainorgcmds]{IEEEtrantools}

\usepackage{bm}

\usepackage{siunitx}
\DeclareSIUnit\atm{atm}

\usepackage[breaklinks=true, linkcolor=blue, citecolor=blue, colorlinks=true]{hyperref}

\usepackage[version=4]{mhchem}

\graphicspath{{figures/}}

\usepackage{xargs}
\usepackage[pdftex,usenames,dvipsnames]{xcolor}

\begin{document}

\articletype{ARTICLE}

\title{Assessing diffusion model impacts on enstrophy and flame 
structure in turbulent lean premixed flames}

\author{
\name{
    Aaron J. Fillo\textsuperscript{a}, 
    Peter E.~Hamlington\textsuperscript{b}, and 
    Kyle E.~Niemeyer\textsuperscript{a}\thanks{CONTACT K.~E.~Niemeyer. Email: kyle.niemeyer@oregonstate.edu}
    }
\affil{
    \textsuperscript{a}School of Mechanical, Industrial, and Manufacturing Engineering, Oregon State University, Corvallis, OR, USA; 
    \textsuperscript{b}Paul M.~Rady Department of Mechanical Engineering, University of Colorado, Boulder, CO 80309, USA
    }
}

\maketitle

\begin{abstract}
Diffusive transport of mass occurs at small scales in turbulent premixed flames. 
As a result, multicomponent mass diffusion, which is often neglected in direct 
numerical simulations (DNS) of premixed combustion, has the potential to impact 
both turbulence and flame characteristics at small scales. 
In this study, we evaluate these impacts by examining enstrophy dynamics and the 
internal structure of the flame for lean premixed hydrogen-air combustion, 
neglecting secondary Soret and Dufour effects.
We performed three-dimensional DNS of these flames by implementing the 
Stefan--Maxwell equations in the code NGA to represent multicomponent mass transport, 
and we simulated statistically planar lean premixed hydrogen-air flames using both 
mixture-averaged and multicomponent models.
The mixture-averaged model underpredicts the peak enstrophy in the multicomponent 
simulation by up to \SI{13}{\percent} in the flame front.
Comparing the enstrophy budgets of these flames, the multicomponent simulation 
yields larger peak magnitudes compared to the mixture-averaged simulation 
in the reaction zone, showing differences of \SI{17}{\percent} and \SI{14}{\percent} 
in the normalized vortex stretching and viscous effects terms.
In the super-adiabatic regions of the flame, the mixture-averaged model overpredicts 
the viscous effects by up to \SI{13}{\percent}. 
To assess the effect of these differences on flame structure, we reconstructed the 
average local internal structure of the turbulent flame through statistical analysis 
of the scalar gradient field. 
Based on this analysis, we show that large differences in viscous effects contribute 
to significant differences in the average local flame structure between the two models.
\end{abstract}


\section{Introduction}\label{Introduction}
The average internal flame structure and statistical behavior of vorticity and enstrophy are 
critical for evaluating the properties and effects of turbulent fluid motions in reacting 
flows~\cite{Hamlington2011,Chakraborty2016,Bobbitt2016}. 
Both chemical heat release and advective mixing by turbulence can form steep gradients in 
temperature and scalar fields, increasing the importance of diffusive transport relative 
to effects due to chemical reactions in the flame dynamics \cite{Bird1960}. 
Prior studies have shown that these enhanced diffusive effects are especially important 
in lean hydrogen flames where the Lewis number is much less than unity, 
leading to an increase in the turbulent flame speed \cite{Lipatnikov2005}. 
A number of computational studies have also examined turbulent premixed flame 
characteristics for lean and low Lewis-number conditions 
\cite{Day:2009,AspdenJFM:2011,Chakraborty2016,Aspden:2017,Schlup2017}. 
Day et al.~\cite{Day:2009}, in particular, showed that the effects of thermo-diffusive 
instabilities remain evident even in turbulent premixed flames where velocity fluctuations 
are nearly three times larger than the laminar flame speed, although stronger turbulence 
moderates the growth of cellular structures associated with these instabilities.

Several studies have already examined the importance of thermal diffusion in a wide range 
of flame configurations~\cite{Coffee:1981,Ern:1998,Ern:1999,Bongers:2003,Day:2009,Yang2010,AspdenJFM:2011,Xin2012,Giovangigli2015MulticomponentFlames,Aspden:2017,Schlup2017}.
These studies have thoroughly demonstrated that neglecting thermal diffusion, in particular, 
can significantly impact flame properties. For example, studying three-dimensional (3D), 
premixed, turbulent hydrogen/air flames, 
Schlup and Blanquart~\cite{Schlup2017} showed that 
thermal diffusion can increase flame propagation speeds and chemical source terms of products 
in regions of high positive curvature.
For laminar hydrogen/air flames, Giovangigli~\cite{Giovangigli2015MulticomponentFlames} 
demonstrated that multicomponent Soret effects can significantly influence laminar 
flame speeds and extinction stretch rates for flat and stretched premixed flames, respectively.
Based on these and other studies, thermal diffusion is thus important for understanding 
and accurately predicting the dynamics of some fuel/air mixtures. 

Although several of the studies noted above have also evaluated the impact of multicomponent 
thermal diffusion models on lean hydrogen flame 
simulations~\cite{Ern:1998,Ern:1999,Dworkin2009TheFlames,Xin:2015,Schlup2017}, 
there is still an incomplete understanding of the impact of full multicomponent mass 
diffusion, a component of the exact scalar (e.g., chemical species concentration or 
mass fraction) governing equations, on turbulent transport and average flame structure.
In many simulation studies of premixed flames, mass diffusion is represented using a 
mixture-averaged approximation, but this may significantly impact the 
coupled dynamics of scalars and turbulence, particularly in lean premixed hydrogen 
flames where thermo-diffusive instabilities are prominent. 
Recently, Fillo et al.~\cite{Fillo2021} studied premixed, turbulent, 
high Karlovitz-number hydrogen/air, \textit{n}-heptane/air, and toluene/air flames,
showing that using the mixture-averaged diffusion model noticeably alters diffusion fluxes
compared with the multicomponent diffusion model.
These variations lead to differences of \SIrange{5}{20}{\percent} in normalized turbulent 
ﬂame speeds and conditional means of fuel source term.
These observations motivate this deeper dive into the impacts of mass diffusion model on turbulence 
and flame dynamics.

The primary objective of this study is to evaluate the impact of the mixture-averaged 
diffusion approximation on enstrophy transport and the average internal flame structure 
using data from direct numerical simulations (DNS) of turbulent, premixed, lean 
hydrogen/air flames. 
This objective will be achieved in two ways.
First, we will analyze the differences in time- and spatially averaged enstrophy budgets 
in the DNS using mixture-averaged and multicomponent diffusion models. 
Second, we will evaluate the impact of these differences on the average local flame 
structure by evaluating scalar gradient trajectories for the two models and statistically 
reconstructing the internal flame structures. 
Based on the results of these analyses, we will assess the accuracy and appropriateness 
of the mixture-averaged diffusion approximation for use in DNS of turbulent, premixed, 
lean hydrogen/air flames.

The paper is organized as follows. Section~\ref{sec:numericalapproach} describes the 
governing equations, diffusion models, and flow configuration for the 3D DNS.
Then, Section~\ref{sec:results} presents a qualitative description of the 
scalar and vorticity magnitude fields, the time- and spatially averaged enstrophy budgets, 
and the statistical reconstruction of the average turbulent flame width. 
Finally, in Section~\ref{sec:conclusion} we draw conclusions from the comparisons of the 
diffusion models.

\section{Numerical approach}
\label{sec:numericalapproach}
This section describes the governing reacting-flow equations, including brief discussions of 
the diffusion models to be studied. We also describe the 3D flow configuration modeled in the 
simulations.

\subsection{Governing equations}
The variable-density, low Mach number, reacting-flow equations are solved using the 
finite-difference code NGA~\cite{Desjardins2008,Savard2015AChemistry}.
The conservation equations are written as
\begin{IEEEeqnarray}{rCl}
\frac{\partial\rho}{\partial t}+\nabla\cdot(\rho\bm{u}) &=& 0 \;, \label{1} \\
\frac{\partial(\rho \bm{u})}{\partial t}+\nabla\cdot(\rho \bm{u}\otimes \bm{u}) &=& -\nabla p+\nabla\cdot\bm{\tau}+\bm{f} \;, \label{2} \\
\frac{\partial(\rho T)}{\partial t}+\nabla\cdot(\rho \bm{u} T) &=& \nabla\cdot \left( \rho\alpha\nabla T \right) - \frac{1}{c_{p}}\sum_{i} c_{p,i}\bm{j}_{i} \cdot \nabla T + \rho \dot{\omega}_{T} + \frac{\rho\alpha}{c_{p}} \nabla c_{p} \cdot \nabla T \;, \label{3} \\
\frac{\partial(\rho Y_{i})}{\partial t}+\nabla\cdot(\rho \bm{u} Y_{i}) &=& -\nabla\cdot \bm{j}_{i}+\rho\dot{\omega_{i}} \;, \label{4}
\end{IEEEeqnarray}
where $\rho$ is the mixture density, $\bm{u}$ is the velocity, $p$ is the hydrodynamic pressure,
$\bm{\tau}$ is the viscous stress tensor, $\bm{f}$ represents volumetric forces, 
$T$ is the temperature, $\alpha$ is the mixture thermal diffusivity, $c_{p,i}$ is the 
constant-pressure specific heat of species $i$, $c_{p}$ is the constant-pressure specific 
heat of the mixture, and $\bm{j}_{i}$, $Y_{i}$ , and $\dot{\omega_{i}}$ are the diffusion 
flux, mass fraction, and production rate of species $i$, respectively. 
In Eq.~\eqref{3}, the temperature source term is given by
\begin{equation}
\dot{\omega}_{T}=-c_{p}^{-1}\sum_{i} h_{i}(T)\dot{\omega_{i}} \;,
\label{5}
\end{equation}
where $h_{i}(T)$ is the specific enthalpy of species $i$ as a function of temperature. 
The density is determined from the ideal gas equation of state.

\subsection{Overview of diffusion models}

The diffusion fluxes are calculated using the semi-implicit scheme developed by 
Fillo et al.~\cite{Fillo2020_jcp} with either the
mixture-averaged \cite{Curtiss1949TransportMixtures,Bird1960} or 
multicomponent \cite{Hirschfelder1954} model, both of which are based on 
Boltzmann's equation for the kinetic theory of 
gases~\cite{Curtiss1949TransportMixtures,Hirschfelder1954}.
We neglect both baro-diffusion and thermal diffusion (i.e., Soret and Dufour effects). 
The baro-diffusion term is commonly neglected in reacting-flow simulations under the 
low-Mach-number approximation~\cite{Grcar2009TheFlames}, and we neglected thermal 
diffusion because our objective is to investigate the impact of mass diffusion models only; 
Schlup and Blanquart~\cite{Schlup2017} previously explored the effects of 
thermal diffusion models on lean premixed flames.

The species diffusion flux for the mixture-averaged diffusion model, 
denoted $\bm{j}_{i}^{\text{MA}}$, is related to the species gradient 
by a Fickian formulation and is expressed as
\begin{equation}\label{6}
\bm{j}_{i}^{\text{MA}}=-\rho D_{i, m}\frac{Y_i}{X_{i}}\nabla X_{i}+\rho Y_{i}\bm{u}_{c} \;,
\end{equation}
where $X_i$ is the $i$th species mole fraction and $D_{i,m}$ is the $i$th species 
mixture-averaged diffusion coefficient.
This was originally introduced by Curtiss and Hirschfelder~\cite{Curtiss1949TransportMixtures}
and is expressed by Bird et al.~\cite{Bird1960}\footnote{Interestingly, the formula for 
mixture-averaged diffusion coefficient is not available in the later editions of 
Bird et al.~\cite{Bird1960}, but it is available in other texts such as that 
by Kee et al.~\cite{KeeCRF}.} as
\begin{equation}\label{7}
D_{i, m}=\frac{1-Y_{i}}{\sum_{i\neq j}^{N} X_{j}/\mathcal{D}_{ji}} \;.
\end{equation}
Here $\mathcal{D}_{ji}$ is the binary diffusion coefficient for species $i$ and $j$, 
and $\bm{u}_{c}$ is the correction velocity used to ensure mass continuity:
\begin{equation} \label{8}
\bm{u}_{c}=\sum_{i} D_{i, m}\frac{Y_{i}}{X_{i}}\nabla X_{i} \;.
\end{equation}
While the mixture-averaged diffusion coefficient and correction velocity were
introduced empirically, Giovangigli~\cite{Giovangigli1991} showed that the resulting 
diffusion flux corresponds to the first term of a series converging towards the exact 
solution of the Stefan--Maxwell equations.

The species diffusion flux for the multicomponent diffusion model, denoted 
$\bm{j}_{i}^{\text{MC}}$, as presented by Bird et al.~\cite{Bird1960} and 
implemented in CHEMKIN II \cite{Kee1989Chemkin-II:Kinetics}, is
\begin{equation} \label{9}
\bm{j}_{i}^{\text{MC}}=\frac{\rho Y_{i}}{X_{i}W}\sum_{i\neq j}^{N}W_{j}D_{i,j}\nabla{X_{j}} \;,
\end{equation}
where $W$ is the mixture molecular weight, $W_{j}$ is the molecular weight of the $j$th species, 
and $D_{i,j}$ is the multicomponent diffusion coefficient computed using the \texttt{MCMDIF} 
subroutine of CHEMKIN II~\cite{Kee1989Chemkin-II:Kinetics} with the method outlined by 
Dixon-Lewis~\cite{Dixon-Lewis1968FlameSystems}.

\begin{table}[htbp]
    \caption{Parameters describing the mixture-averaged (MA) and multicomponent (MC) 
    diffusion simulations, with definitions provided, where $u'$ is the turbulent 
    fluctuating velocity, $\epsilon$ is the turbulent energy dissipation rate, and 
    $\nu_\text{u}$ is the unburnt kinematic viscosity.}
    \centering
    \footnotesize{
    \begin{tabular}{@{}l c c l@{}}
         \toprule
         & MA & MC & Description \\
         \midrule
         Domain & \multicolumn{2}{c}{$8L \times L \times L$} & Dimensions of the computational domain \\
         $L$ & \multicolumn{2}{c}{190$\Delta{x}$} & Spanwise width of the computational domain\\
         Grid & \multicolumn{2}{c}{$1520\times190\times190$} & Computational grid size \\
         $\Delta{x}$ [\si{\mm}] & \multicolumn{2}{c}{0.0424} & Computational grid spacing \\
         $\eta_\text{u}$ [\si{\m}] & \multicolumn{2}{c}{\num{2.1e-5}} & Kolmogorov length scale in the unburnt gas \\
         $\tau_\eta$ [\si{\s}] & \multicolumn{2}{c}{\num{1.87e-5}} & Kolmogorov time scale in the unburnt gas \\
         $\Delta{t}$ [\si{\s}] & \multicolumn{2}{c}{\num{6e-7}} & Simulation time-step size \\
         $\phi$ & \multicolumn{2}{c}{0.4} & Equivalence ratio \\
         $P_0$ [\si{\atm}] & \multicolumn{2}{c}{\num{1}} & Pressure of the unburnt mixture\\
         $T_{\text{u}}$ [\si{\K}] & \multicolumn{2}{c}{\num{298}} & Temperature of the unburnt mixture \\
         $T_{\text{peak}}$ [\si{\K}] & 1190 & 1180 & Temperature of peak fuel consumption rate in a laminar flame\\
         $T_{\text{b}}$ [\si{\K}] & 1422 & 1422 & Temperature of the burnt mixture in a laminar flame \\
         $S_L$ [\si{\m\per\s}] & 0.230 & 0.223 & Laminar flame speed \\
         $\delta_L$ [\si{\mm}] & 0.643 & 0.631 & Laminar flame thermal thickness $\delta_L = \left(T_\text{b} - T_\text{u}\right)/\left|\nabla T\right|_{\max}$\\
         $\tau_L$ [\si{\s}] & \num{2.80e-3} & \num{2.83e-3} & Laminar flame characteristic time scale $\tau_L = \delta_L/S_L$\\
         $\ell/\delta_L$ & 2.00 & 2.04 & Integral length scale $\ell = u'^3/\epsilon$ relative to $\delta_L$\\
         $u'/S_L$ & 18.0 & 18.6 & Turbulent fluctuation velocity relative to $S_L$ \\
         $\text{Ka}_\text{u}$ & 149 & 151 & Karlovitz number in the unburnt mixture, $\text{Ka}_\text{u} = \tau_{L}/\tau_{\eta}$ \\
         $\text{Re}_\text{t}$ & \multicolumn{2}{c}{289} & Reynolds number in the unburnt mixture, $\text{Re}_\text{t} = (u'l)/\nu_\text{u}$ \\
         \bottomrule
    \end{tabular}}
    \label{tab:3D_flow_config}
\end{table}

\subsection{Simulation configuration}
The simulations model a 3D statistically stationary, statistically planar lean premixed 
hydrogen/air flame \cite{Burali2016AssessmentFlows,Lapointe2016FuelFlames,Schlup2017}. 
This fuel/air mixture has a low Lewis number ($\mathrm{Le}_{\rm H_2} = 0.3$) and was 
selected because the fidelity of the diffusion model (i.e., mixture-averaged or multicomponent) 
may be important for accurately simulating the instabilities associated with differential 
diffusion in lean hydrogen/air flames. Chemical reactions in the hydrogen/air mixture are 
represented using the nine-species, 54-reaction chemistry model from 
Hong et al.~\cite{Hong2011AnMeasurements,Lam2013AAbsorption,Hong2013OnAbsorption} 
(forward and backward reactions are counted separately).

The 3D turbulent flames are simulated using an identical flow configuration as in previous 
studies \cite{Savard2015,Burali2016AssessmentFlows,Schlup2017,Fillo2020_jcp,Fillo2021}, and 
therefore we only briefly describe them here. 
The computational domain consists of inflow and convective outflow boundary conditions in the 
stream-wise (i.e., $x$) direction. 
The two span-wise directions (i.e., $y$ and $z$) use periodic boundaries. 
The inflow velocity is the mean turbulent flame speed, which keeps the flame statistically 
stationary such that turbulent statistics can be collected over an arbitrarily long run time. 
In the absence of mean shear, a linear turbulence forcing method \cite{Rosales2005,Carroll2013} 
is implemented to maintain the production of turbulent kinetic energy through the flame. 
Klein et al.~\cite{Klein2017} showed that, in statistically planar flame 
configurations such as those examined here, neither unforced decaying turbulence, 
boundary-only forcing, or linear forcing are clearly preferable, indicating that the 
linear forcing used here is sufficient for examining the relative impacts of different 
diffusion models in the present simulations.

Table~\ref{tab:3D_flow_config} provides further details of the computational domain, 
unburnt mixture, corresponding one-dimensional flame statistics, and inlet turbulence 
in both the mixture-averaged (MA) and multicomponent (MC) diffusion simulations. 
The unburnt temperatures and pressures are \SI{298}{\kelvin} and \SI{1}{\atm}, respectively. 
The definitions of the unburnt Karlovitz number, $\text{Ka}_\text{u}$, and turbulent Reynolds 
number, $\text{Re}_\text{t}$, are also given in Table~\ref{tab:3D_flow_config}, where 
$\tau_L = \delta_L / S_L$ is the flame time scale and 
$\tau_{\eta}=\left(\nu_\text{u} / \epsilon\right)^{1/2}$ is the Kolmogorov time scale 
of the incoming turbulence. 
The forcing was designed to produce a turbulence integral scale $\ell=u'^3/\varepsilon$ 
of roughly $2\delta_L$; directly calculating the integral scale from the longitudinal velocity
correlation after the simulations were performed yields $\ell\approx 2.3\delta_L$, 
close to the intended integral scale.
Based on this computed integral scale, the position of the flame in the simulation domain is 
roughly $8\ell$ from the inlet at $x/L=0$, suggesting that results in the flame region should 
not be strongly affected by the inlet boundary condition.
We modeled relatively high Karlovitz numbers of $\mathrm{Ka}_{\text{u}} \approx 150$ in both 
simulations to represent cases where the turbulence timescale is shorter than the 
species diffusion timescale, resulting in pronounced impacts of the turbulence on the flame. 

\section{Results and discussion}\label{sec:results}

In this section, we first present a qualitative description of the instantaneous velocity and 
scalar fields. Next, we present a time- and spatially averaged assessment of the enstrophy 
budget, followed by a statistical reconstruction of the average local flame structure.

\subsection{Qualitative description}
\label{sec:qualitative}

As an initial assessment of impact of mixture-averaged and multicomponent mass diffusion on 
flame dynamics, we present instantaneous flow fields for the simulated flames. 
Both simulations are initialized with the same scalar and velocity fields, and run for a 
single time iteration to evaluate the impact of diffusion on the scalar field, 
independent of turbulent mixing.

\begin{figure}[tbp]
  \includegraphics[width=\textwidth]{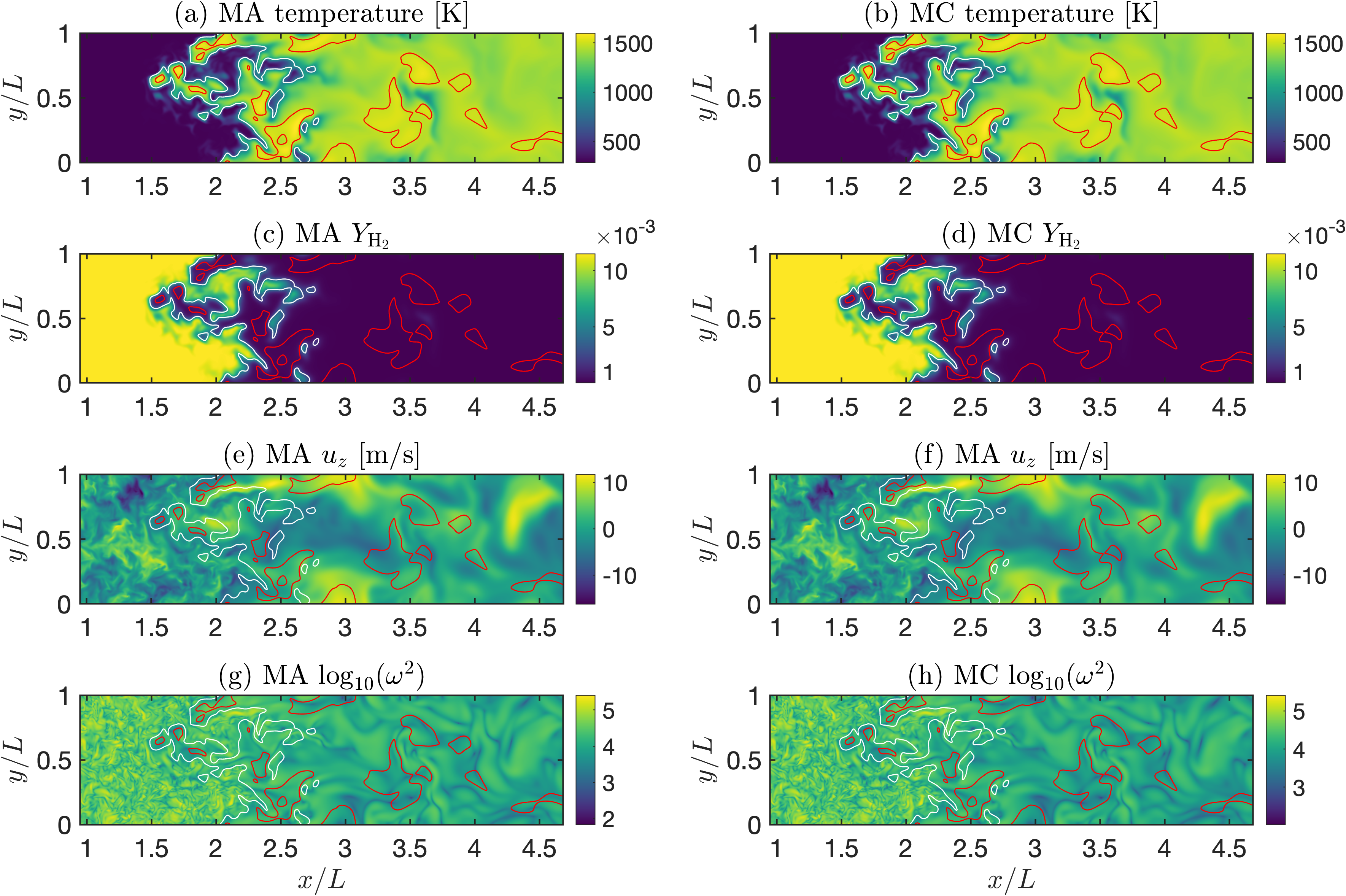}
  \caption{Instantaneous fields of temperature $T$ (a, b), 
  fuel mass fraction $Y_{\text{H}_2}$ (c, d), $z$-direction velocity $u_z$ (e, f), 
  and vorticity magnitude $\omega^2$ (g, h) for one time step of the hydrogen/air 
  turbulent premixed flame for the mixture-averaged (MA) and multicomponent (MC) diffusion cases. 
  Shown are domain cross-sections through the midplane. 
  The red and white lines correspond to isosurfaces of $T_0=T_{\text{peak}}-\SI{300}{\kelvin}$ 
  and $T_f=T_{\text{peak}}+\SI{300}{\kelvin}$, respectively, and represent the inflow and 
  outflow surfaces of the flame front.}
  \label{fig:contours}
\end{figure}

Figure~\ref{fig:contours} shows contours of temperature, hydrogen mass fraction, $z$-direction 
velocity, and the logarithm of the total vorticity magnitude 
( $\omega=(\bm{\omega}\cdot \bm{\omega})^{1/2}$), where 
$\bm{\omega}=\nabla \times \bm{u}$. 
The inlet and outlet of the flame front are defined by the isosurfaces, 
$T_0=T_{\text{peak}}-\SI{300}{\kelvin}$ and $T_f=T_{\text{peak}}+\SI{300}{\kelvin}$, 
respectively, where $T_{\text{peak}}$ is the temperature of peak fuel consumption rate in 
the one-dimensional laminar flame (see Table~\ref{tab:3D_flow_config}). 
Figures~\ref{fig:contours}(a,b) and (c,d) show that temperature increases across the flame 
as the fuel is consumed in both simulation cases.
Figures~\ref{fig:contours}(e,f) show that the velocity fields $u_z$ become smoother with 
fewer small-scale features across the flame, corresponding to an overall reduction in the 
vorticity magnitude from reactants to products, as Figs.~\ref{fig:contours}(g,h) show.

Shown qualitatively in Fig.~\ref{fig:contours}, the $T_f$ isosurface is located at the 
approximate transition point between the preheat and reaction zones, while the $T_0$ 
isosurface captures the super-adiabatic regions, also called ``hot spots'', present in 
lean premixed hydrogen flames.
These hot spots result from differential diffusion and have been predicted by
theory~\cite{Williams:1985} and shown in simulations of lean, premixed hydrogen/air 
flames, including in the post-flame 
region~\cite{Day:2009,AspdenJFM:2011,Aspden:2015,Aspden:2017}.
Fillo et al.~\cite{Fillo2021} previously showed that the two diffusion models result in 
differences of \SIrange{5}{35}{\percent} in conditional means of both fuel mass fraction 
and source term in these hot spots.

\begin{figure}[htbp]
\centering
\includegraphics[width=0.8\textwidth]{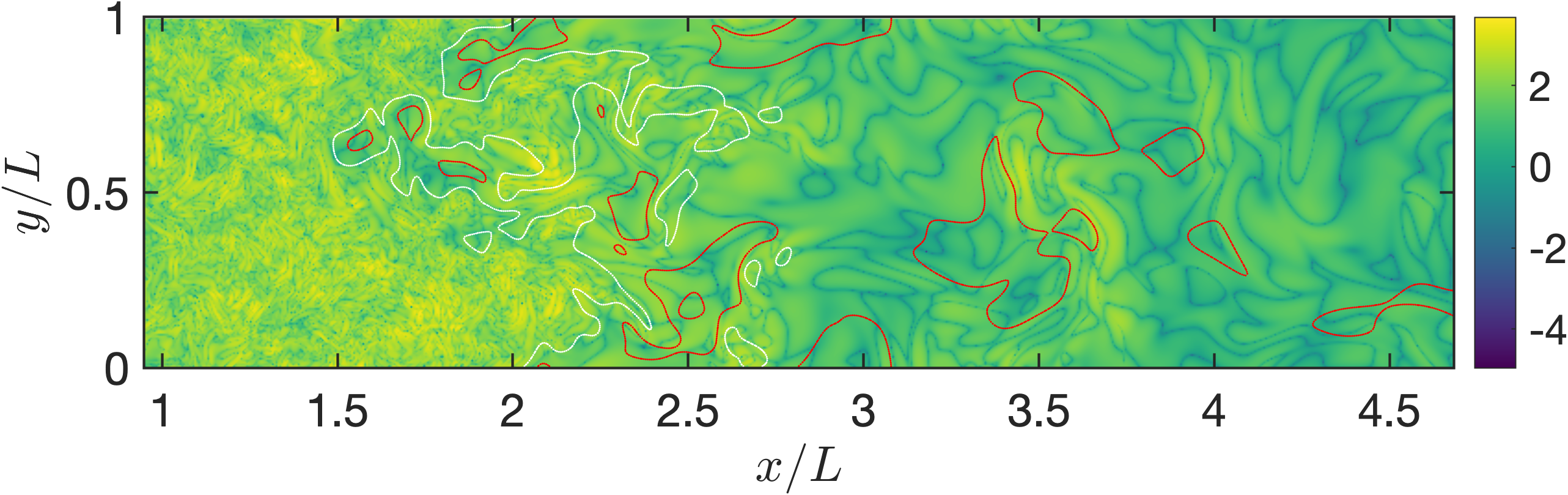}
\caption{Instantaneous field of the difference in vorticity magnitude, 
$\log_{10}(|\omega_{\text{MA}}^2-\omega_{\text{MC}}^2|)$, between the mixture-averaged and multicomponent diffusion models for the lean premixed hydrogen-air flame examined here. Shown is the domain cross-section through the midplane. The red and white lines correspond to isosurfaces of $T_0=T_{\text{peak}}-\SI{300}{\kelvin}$ and $T_f=T_{\text{peak}}+\SI{300}{\kelvin}$, respectively, representing the inflow and outflow surfaces of the flame front.} 
\label{fig:Vort_diff}
\end{figure}

As expected, the mixture-averaged and multicomponent contours in Fig.~\ref{fig:contours} 
exhibit little difference for a single time step.
However, as shown in Fig.~\ref{fig:Vort_diff}, if we examine the difference in the vorticity 
magnitude as an indicator of the relative impact of diffusion model on turbulent transport 
through the flame, the two cases notably disagree even after only a single time step.
Although qualitative, Fig.~\ref{fig:Vort_diff} highlights the impact that diffusion model 
can have on the turbulent flow, thereby impacting turbulence-flame interactions at these 
high-Karlovitz conditions.
On average, these differences can result in a significant and measurable difference in global 
flame statistics.


\subsection{Enstrophy dynamics}

To begin the analysis of enstrophy dynamics, we first consider time- and spatially averaged 
enstrophy ($\omega^2$) for the mixture-averaged and multicomponent simulations. 
We computed these statistics over 25 eddy turnover times ($\tau$), where 
$\tau=k/\epsilon\approx$ \SI{500}{\micro\second}, after first allowing the flames to 
develop in a turbulent flow field (ensuring that all transients from the initialization are lost). 
The spatial averages are calculated in $y$-$z$ spanwise planes along the $x$ direction.

Figure~\ref{fig:enstrophy} shows that the small differences that appear in one time step 
(see Fig.~\ref{fig:Vort_diff}) grow in magnitude over the course of the simulation. 
In particular, the mixture-averaged model underpredicts the peak enstrophy of the 
multicomponent diffusion model by up to \SI{13}{\percent} in the flame front.
This difference suggests that the intensity of small-scale turbulence is generally 
lower in the mixture-averaged case, as compared to the corresponding multicomponent case. 
Moreover, the 13\% quantitative difference in peak enstrophy shown in 
Fig.~\ref{fig:enstrophy} closely matches the difference in mean turbulent flame speed 
in these flames shown previously by Fillo et al.~\cite{Fillo2021}. 

However, for values of $x/\delta_L$ greater than roughly 3, 
Fig.~\ref{fig:enstrophy} shows that the enstrophy in the mixture-averaged case is actually 
greater than in the multicomponent case. 
This corresponds to the super-adiabatic regions of the flame, 
{discussed in Section~\ref{sec:qualitative},
and indicates that the choice of mass diffusion model can have different relative effects on 
the local enstrophy magnitude at different locations in the high-Karlovitz, low-Lewis number 
premixed flames considered here. 

\begin{figure}[htbp]
    \centering
    \includegraphics[width=0.7\textwidth]{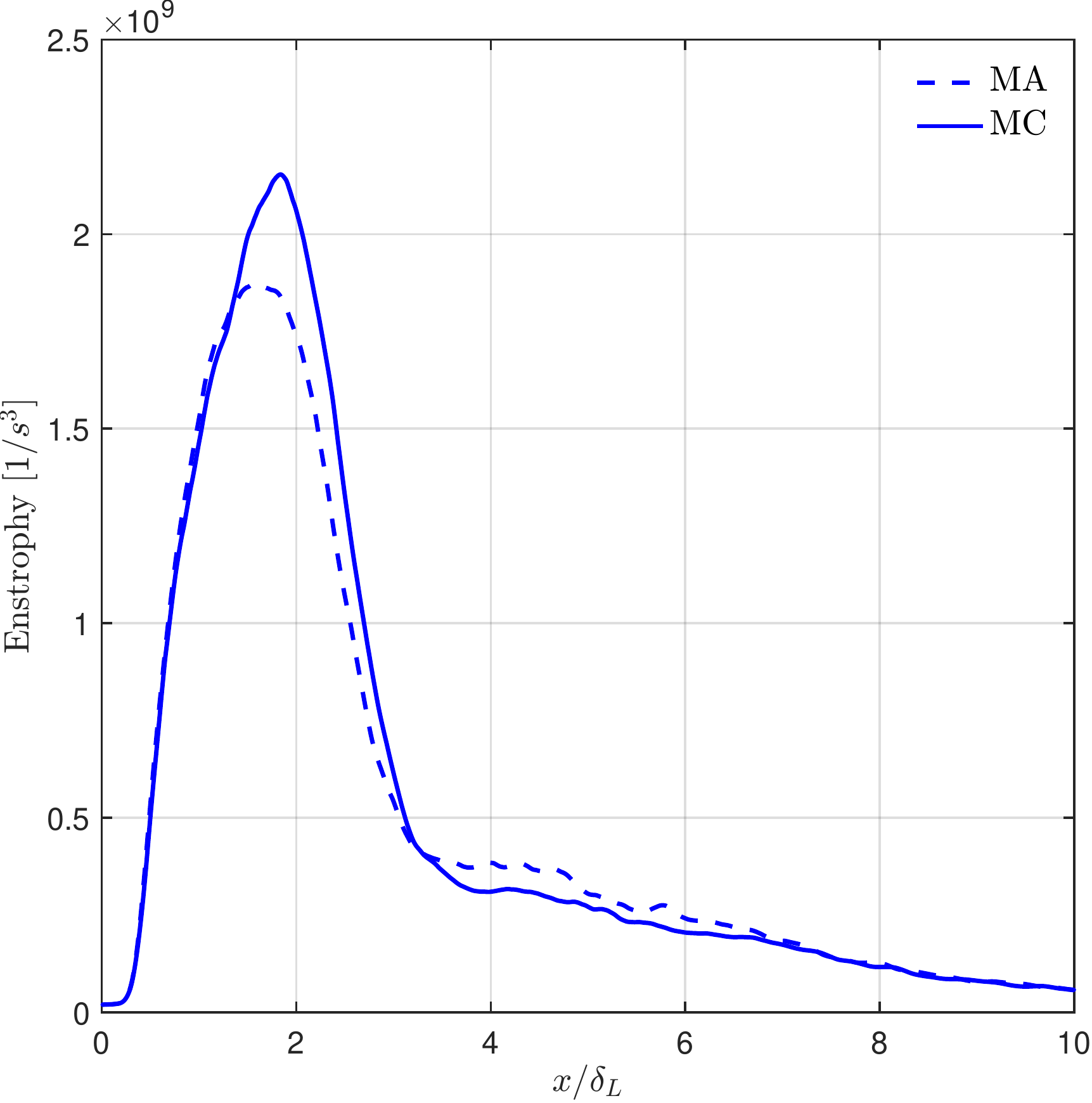}
    \caption{Time and spatially (spanwise) averaged enstrophy for multicomponent (MC) and 
    mixture-averaged (MA) diffusion models.}
    \label{fig:enstrophy}
\end{figure}

To explain the differences in the vorticity magnitude\slash enstrophy when using 
mixture-averaged and multicomponent mass diffusion models, next we examine the transport 
equation for the enstrophy, which is obtained from the curl of the momentum equation in 
Eq.~\eqref{2} as
{\begin{IEEEeqnarray}{rCl}
    \frac{1}{2}\frac{D\omega^2}{Dt} = \underbrace{\bm{\omega}\cdot(\bm{\omega}\cdot\nabla)\bm{u}}_{\mathrm{Stretching}} - \underbrace{\omega^2(\nabla\cdot\bm{u})}_{\mathrm{Dilatation}} + \underbrace{\frac{\bm{\omega}}{\rho^2}\cdot(\nabla\rho\times\nabla{p})}_{\mathrm{Baroclinic~torque}} + \underbrace{\bm{\omega}\cdot\nabla\times\left(\frac{1}{\rho}\nabla\cdot\bm{\tau}\right)}_{\mathrm{Viscous~effects}} + \underbrace{\bm{\omega}\cdot\nabla\times\frac{\bm{f}}{\rho}}_{\mathrm{Forcing}} \;, \IEEEeqnarraynumspace
    \label{eqn:enstrophy}
\end{IEEEeqnarray}}%
where $D/Dt=\partial/\partial t+\bm{u}\cdot \nabla$ is the material derivative.
The first term on the right-hand side of Eq.~\eqref{eqn:enstrophy} represents vortex stretching, 
which is a nonlinear term accounting for the interaction between the strain rate and vorticity.
The second term represents dilatation, which is primarily negative in reacting flows, 
suppressing vorticity magnitude. 
The third term represents baroclinic torque, which is only substantially nonzero when the 
gradients of density and pressure are both nonzero and misaligned. 
The fourth term represents viscous effects and includes contributions from both viscous 
diffusion and viscous dissipation. 
Finally, the last term represents the effect of the numerical body force present in Eq.~\eqref{2}.

Using data from both the mixture-averaged and multicomponent diffusion simulations, 
we compute time and spanwise averages of each term in Eq.~\eqref{eqn:enstrophy} 
to examine the dynamical causes of the different enstrophy magnitudes shown in 
Figs.~\ref{fig:Vort_diff} and \ref{fig:enstrophy}. 
Figure~\ref{fig:enstrophy_budget} shows the resulting normalized time- and spatially 
averaged enstrophy budgets, where we normalized following the approach by 
Bobbitt et al.~\cite{Bobbitt2016}. 

Viscous effects are the primary sink term for both simulation cases and the primary source 
terms are vortex stretching, followed by forcing.
For each of these physical effects, the multicomponent simulation yields larger peak 
magnitudes compared to the mixture-averaged simulation near the reaction zone of the flame 
(corresponding to the region close to $x/\delta_L\approx 2$).
Interestingly, despite underpredicting the peak magnitudes of stretching, viscous, 
and forcing effects in the reaction zone, Fig.~\ref{fig:enstrophy_budget} shows that the 
mixture-averaged model overpredicts the viscous effects by as much as \SI{13}{\percent} 
in the super-adiabatic region of the flame (for values of $x/\delta_L$ greater than roughly 3).
This difference between the two diffusion models mirrors the change in enstrophy within 
the super-adiabatic region shown in Fig.~\ref{fig:enstrophy}, and indicates that turbulence 
may thicken the flame in these super-adiabatic regions; this will be examined in more detail 
in the next section.

Finally, Fig.~\ref{fig:enstrophy_budget} shows that baroclinic torque is weakly positive 
and dilatation is weakly negative between $x/\delta_L\approx 2$--3 for both simulation cases, 
agreeing with prior studies of enstrophy dynamics in highly turbulent statistically planar 
premixed flames (see Steinberg et al.~\cite{Steinberg2021} for a review).
Baroclinic torque can become the dominant term in the 
enstrophy dynamics when there is a persistent background pressure gradient, as in the 
tailored channel bluff body experiments by Geikie and Ahmed~\cite{Geikie2018} and the 
swirl burner experiments by Kazbekov et al.~\cite{Kazbekov2019,Kazbekov2021}. 
However, in unconfined statistically planar premixed flames such as those examined here, 
baroclinic torque becomes increasingly weak compared to stretching and viscous effects as 
the turbulence intensity increases \cite{Bobbitt2016}.
The present results further show that, despite the differences between the multicomponent 
and mixture-averaged cases for the vortex stretching, viscous, and forcing terms, 
the baroclinic torque and dilatation terms are relatively unaffected by the choice of 
mass diffusion model.

\begin{figure}[htbp]
    \centering
    \includegraphics[width=0.7\textwidth]{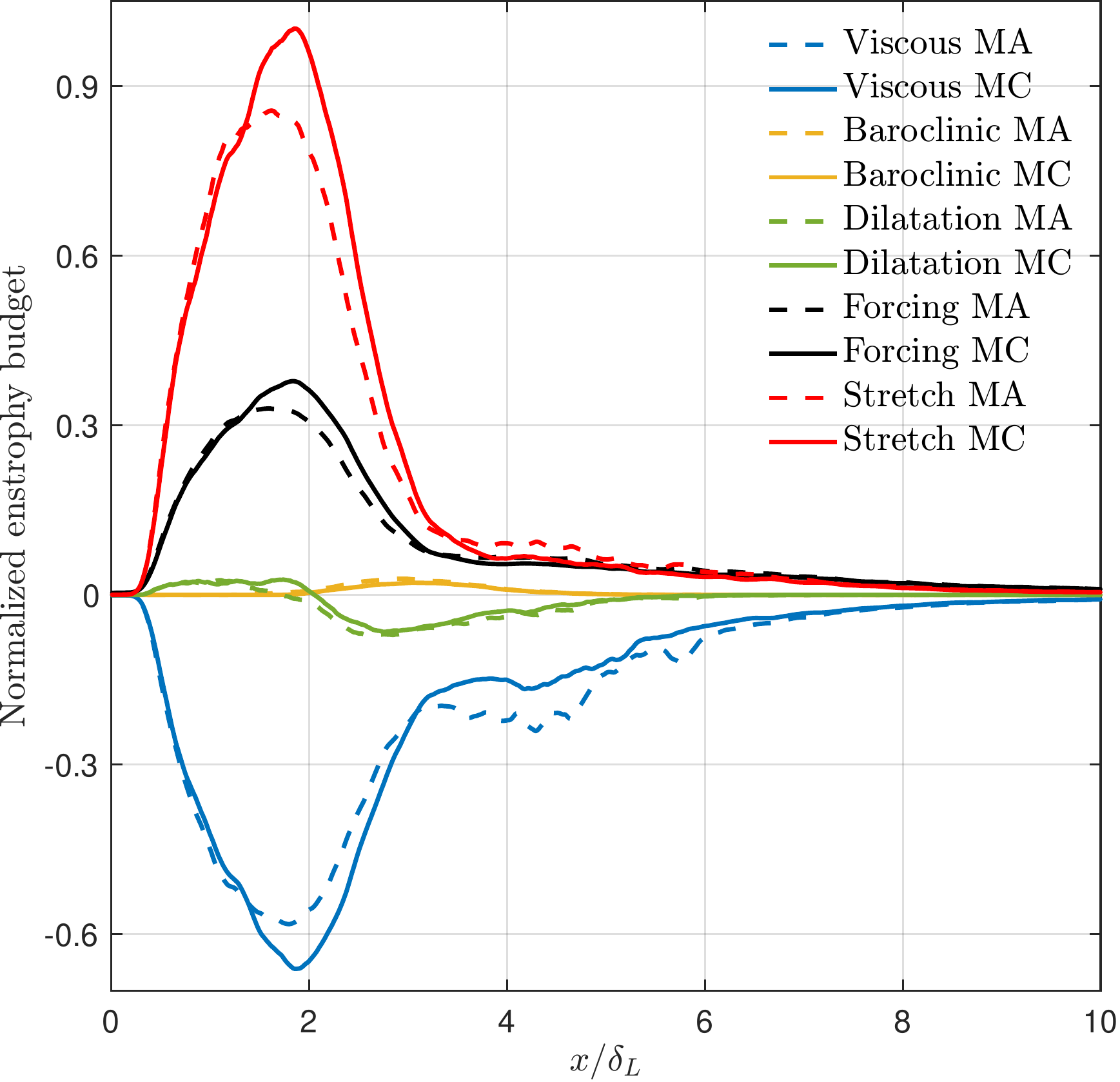}
    \caption{Time and spatially (spanwise) averaged enstrophy budgets for multicomponent (MC) 
    and mixture-averaged (MA) diffusion models.}
    \label{fig:enstrophy_budget}
\end{figure}


\subsection{Flame width and reconstruction}

To evaluate the impact of the observed differences in turbulence dynamics on the global flame 
structure, we reconstructed the average local internal structure of the turbulent flames.
The reconstruction method used here was previously described by 
Hamlington et al.~\cite{Hamlington2011}, and we refer the reader to that study for details.
Briefly, the internal structure of the flame is connected to the magnitude of the temperature 
gradient, $\tilde\chi=(\nabla T \cdot \nabla T)^{1/2}/(T_b-T_u)$, where $T_b$ and 
$T_u$ are the temperatures of the burnt and unburnt mixtures, respectively, in the corresponding 
laminar flame (see Table \ref{tab:3D_flow_config}). 
Large $\tilde{\chi}$ indicates a thin flame and small $\tilde{\chi}$ indicates a broad 
flame \cite{Hamlington2011,Kim2007}. 
Correspondingly, we define $\delta_{t}=\tilde{\chi}^{-1}$ as the local turbulent flame width. 

\begin{figure}[htbp]
    \centering
    \includegraphics[width=0.7\textwidth]{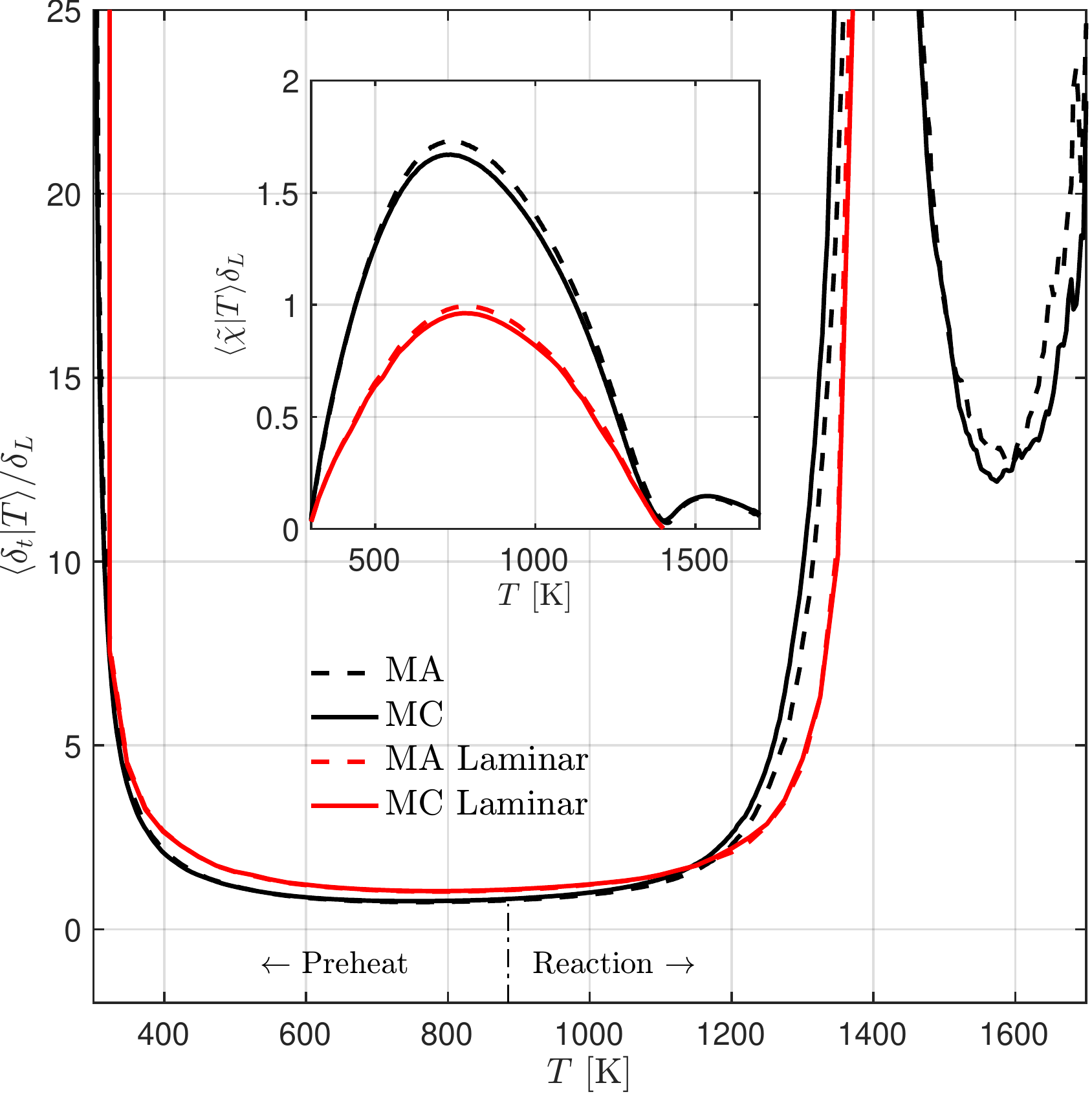}
    \caption{Conditional means of the local flame width $\langle\delta_{t}|T\rangle	\equiv\langle \tilde{\chi}^{-1}|T\rangle$ normalized using the laminar flame thickness, $\delta_{L}$. The inset shows $\langle \tilde{\chi}|T\rangle$ normalized by $\delta_{L}$.} 
    \label{invers_grad}
 \end{figure}
 
\begin{figure}[htbp]
    \centering
    \includegraphics[width=0.7\textwidth]{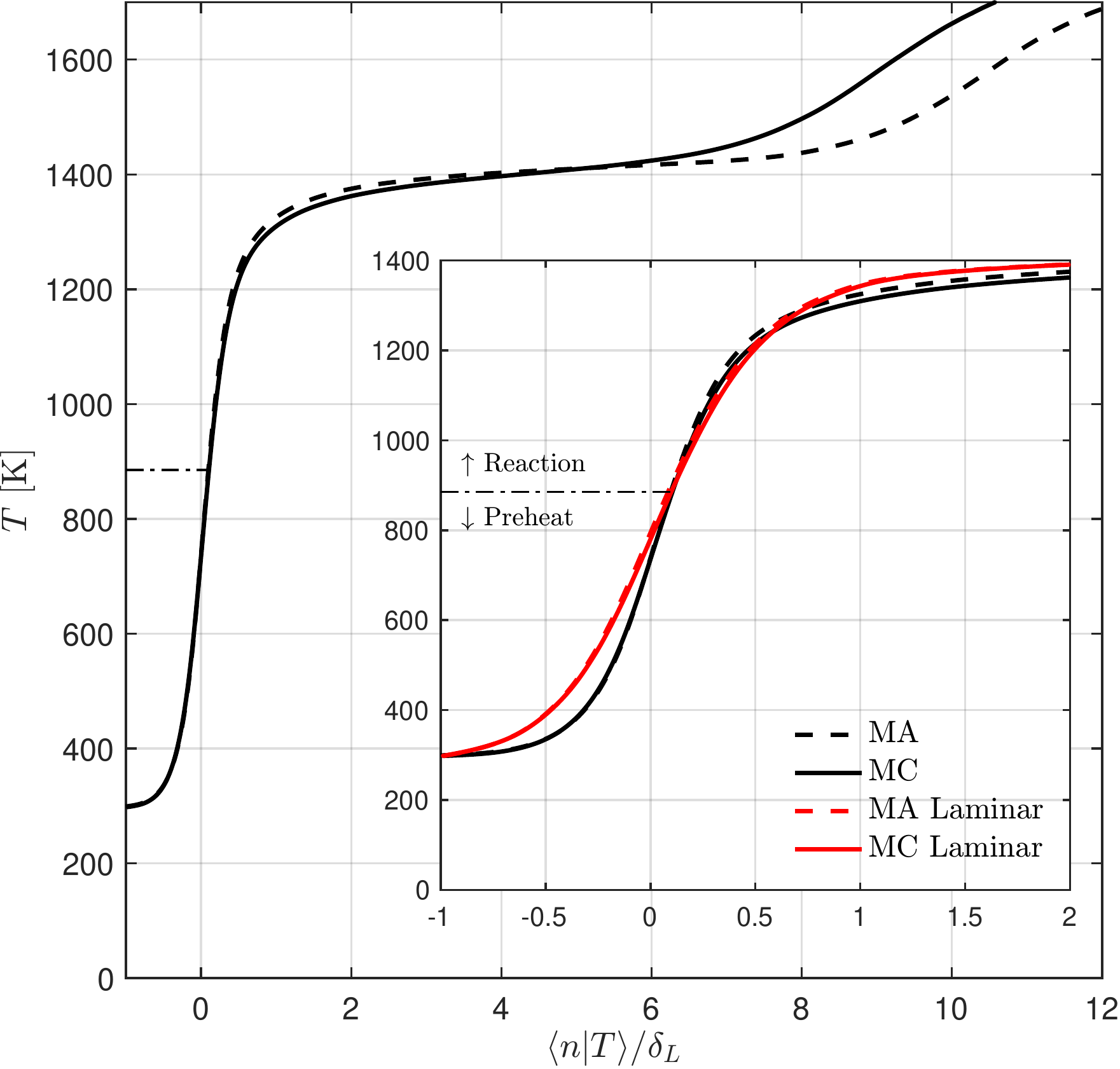}
    \caption{Average local flame structure reconstructed using 
    $\langle \tilde{\chi}^{-1}|T\rangle$ from Fig.~\ref{invers_grad} and 
    Eq.~\eqref{flame_struc} for the turbulent and laminar flames with multicomponent 
    and mixture-averaged mass diffusion. We chose $\langle n|T=T_\mathrm{ref}\rangle=0$ in 
    Eq.~\eqref{flame_struc} by requiring that $\langle n|T_\mathrm{ref}\rangle/\delta_{L}=0$ 
    for all cases, where $T_\mathrm{ref}=T_0$ and $\langle n|T\rangle/\delta_{L}>0$ are 
    locations closer to reactants and $\langle n|T\rangle/\delta_{L}<0$ are locations 
    closer to products. The inset highlights the flame front to facilitate comparison 
    with the average local laminar flame structure.}
    \label{Flame_reconstruct}
\end{figure}

Figure~\ref{invers_grad} shows that for both the  mixture-averaged and multicomponent models, 
the presence of turbulence thins the flame overall, which is expected in the thin-flame regime.
Consistent with the contours shown in Figs.~\ref{fig:contours} and \ref{fig:Vort_diff}, 
we define the separation between the preheat and reaction zones based on the 
$T_0=T_\mathrm{peak}-\SI{300}{\kelvin}$ isosurface. 
Both flames have similar widths in the preheat zone while the multicomponent flame is 
slightly thinner in the reaction zone.
The value of $\langle\delta_{t}|T\rangle/\delta_{L}$ in Fig.~\ref{invers_grad} has a second 
minimum at $\sim$\SI{1600}{\kelvin}, corresponding to the super-adiabatic region of the flame.
At these high temperatures, the mixture-averaged flame is broader, consistent with the observed 
differences in viscous effects in Fig.~\ref{fig:enstrophy_budget}. 
Values of $\langle\delta_{t}|T\rangle/\delta_{L}$ can be greater than one since 
$\delta_{L}\sim |\nabla T|^{-1}_\mathrm{max}$ corresponds to the minimum local 
width of the laminar flame, and both the turbulent and laminar widths exceed 
this at most locations.

Using the distributions of $\langle\delta_{t}|T\rangle$ in Fig.~\ref{invers_grad}, 
we reconstruct the average local internal structure of the turbulent flames using the 
procedure outlined by Hamlington et al.~\cite{Hamlington2011}.
The average flame-normal coordinate, $\langle n|T\rangle$, is calculated from 
$\langle\tilde{\chi}^{-1}|T\rangle/\delta_{L}$ as
\begin{equation}\label{flame_struc}
    \langle n|T\rangle=\langle n|T = T_\mathrm{ref}\rangle + \int_{0}^{T}\langle\tilde{\chi}^{-1}|\eta\rangle d\eta\;,
\end{equation}
where $\langle n|T = T_\mathrm{ref}\rangle$ is the location corresponding to 
$T_\mathrm{ref}=T_0$, taken as the transition between the preheat and reaction zones for 
the present flames.
Integrating Eq.~\eqref{flame_struc} gives profiles of $T$ as a function of $\langle n|T\rangle$,
which approximate the internal structure of the turbulent flame. 

The resulting profiles in Fig.~\ref{Flame_reconstruct} show that the preheat zone thins for 
the turbulent flames and confirms that the multicomponent flame is slightly broader in the 
reaction zone.
Moreover, Fig.~\ref{Flame_reconstruct} shows that the super-adiabatic regions of the 
mixture-averaged flame are as much as \SI{18}{\percent} broader than in the multicomponent flame.
This large difference in flame structure indicates that mixture-averaged diffusion may not 
fully capture the complex interaction between diffusion and turbulent transport in high-temperature 
regions of the flame where steep gradients in the scalar field are present.

These differences may be explained by considering the underlying mathematical forms of the two 
diffusion models.
As shown in Eq.~\eqref{6}, the direction of the mixture-averaged diffusion flux vector is 
strictly aligned counter to the species gradient vector.
Thus, species can only diffuse down the species gradient from high concentration to low 
concentration. Alternatively, as shown in Eq.~\eqref{9}, the multicomponent diffusion model
does not restrict the direction of the diffusion flux vector.
In this case, diffusion can occur in multiple directions simultaneously, 
corresponding to the full scalar field.

Physically, this means that any small-scale changes in the mixture-averaged velocity 
field---on the order of viscous effects---must also be limited in their direction.
By limiting the direction of diffusion, the magnitude of mass transport must increase to 
account for any diffusion not aligned with the species gradient in the multicomponent case.
Aggregating diffusion flux into a single direction may change the direction of local 
velocity, redistributing mass and momentum.

\section{Conclusions} \label{sec:conclusion}

In this study, we assessed the impact of mixture-averaged and multicomponent species 
diffusion models on turbulent enstrophy dynamics and average local flame structure for 3D, 
premixed, high-Karlovitz, lean hydrogen/air flames. 
We observed small differences when comparing the total vorticity magnitude for the two flames 
even after one time step, suggesting that the mixture-averaged diffusion assumption may not fully 
model the physical mass transport of the full multicomponent case.
These differences grow over time, leading to a difference in enstrophy of up to 
\SI{13}{\percent} in the flame front.

Additional time- and spatially averaged analyses of enstrophy transport show 
significant differences in vortex stretching and viscous effects between the two models. 
Specifically, using the mixture-averaged model underpredicts the viscous effects and vortex 
stretching terms of the enstrophy budget by \SI{13}{\percent} and \SI{17}{\percent}, 
respectively, in the flame front.

Variations in the vortex structure and viscous terms reappear in the super-adiabatic 
regions of the flame.
These differences seem to contribute to significant broadening of the mixture-averaged flame 
relative to the multicomponent flame in these regions. 
Thus, although the mixture-averaged diffusion model may adequately reproduce full multicomponent 
mass diffusion in the preheat and reaction zones, it may fail to appropriately model mass 
transport in high-temperature, thermally unstable regions of the flame. 
The strict alignment of the mixture-averaged diffusion vector with the species counter-gradient 
may increase the local velocity and steepen velocity gradients, corresponding 
to an overprediction in the viscous dissipation.

These results suggest that although the mixture-averaged 
diffusion model may reasonably approximate full multicomponent diffusion, 
care should be taken in applying it, particularly in high-Karlovitz, low-Lewis number flames 
similar to those examined here.

Additional study is needed to determine whether stoichiometric or rich premixed flames 
show corresponding effects, and whether diffusion flames are similarly affected. 
Previously, Fillo et al.~\cite{Fillo2021} found differences in the characteristics of nearly 
stoichiometric hydrocarbon flames when using mixture-averaged and multicomponent diffusion 
models, and corresponding differences are likely present in the enstrophy dynamics.
Burali et al.~\cite{Burali2016AssessmentFlows} showed that using a set of constant non-unity 
Lewis numbers produces only small errors in ethylene/air diffusion flames,
as compared with mixture-averaged simulations, suggesting that there may be only slight
differences between mixture-averaged and multicomponent models in diffusion flames.
Further studies should be performed in the future to determine the relative impacts of 
mixture-averaged and multicomponent diffusion models on enstrophy dynamics for a broader 
range of configurations and combustion conditions.

\section*{Acknowledgments}
We thank Clara Llebot Lorente for helping archive the simulation output data.

\section*{Funding}
This material is based upon work supported by the National Science Foundation under 
Grant Nos.\ 1314109-DGE and 1761683.
KEN also acknowledges support from the Welty Faculty Fellowship at Oregon State University.
PEH was supported, in part, by AFOSR Award No.~FA9550–17-1–0144 and NSF Award No.~1847111.
This research used resources of the National Energy Research Scientific Computing 
Center, a DOE Office of Science User Facility supported by the Office of Science
of the U.S.\ Department of Energy under Contract No.\ DE-AC02-05CH11231.

\bibliographystyle{tfq}
\bibliography{references.bib}

\begin{thebibliography}{10}
\newcommand{\printfirst}[2]{#1}
\newcommand{\switchargs}[2]{#2#1}
\providecommand{\url}[1]{\normalfont{#1}}
\providecommand{\urlprefix}{Available at }

\bibitem{Hamlington2011}
P.E. Hamlington, A.Y. Poludnenko, and E.S. Oran, \emph{Interactions between
  turbulence and flames in premixed reacting flows}, Phys. Fluids 23 (2011), p.
  125111, \url{https://doi.org/10.1063/1.3671736}.

\bibitem{Chakraborty2016}
N. Chakraborty, I. Konstantinou, and A. Lipatnikov, \emph{Effects of {Lewis}
  number on vorticity and enstrophy transport in turbulent premixed flames},
  Phys. Fluids 28 (2016), p. 015109, \url{https://doi.org/10.1063/1.4939795}.

\bibitem{Bobbitt2016}
B. Bobbitt, S. Lapointe, and G. Blanquart, \emph{Vorticity transformation in
  high {Karlovitz} number premixed flames}, Phys. Fluids 28 (2016), p. 015101,
  \url{https://doi.org/10.1063/1.4937947}.

\bibitem{Bird1960}
R.B. Bird, W.E. Stewart, and E.N. Lightfoot, \emph{{Transport Phenomena}}, John
  Wiley {\&} Sons, Inc., New York, 1960.

\bibitem{Lipatnikov2005}
A.N. Lipatnikov and J. Chomiak, \emph{Molecular transport effects on turbulent
  flame propagation and structure}, Progress in Energy and Combustion Science
  31 (2005), pp. 1--73, \url{https://doi.org/10.1016/j.pecs.2004.07.001}.

\bibitem{Day:2009}
M. Day, J. Bell, P.T. Bremer, V. Pascucci, V. Beckner, and M. Lijewski,
  \emph{Turbulence effects on cellular burning structures in lean premixed
  hydrogen flames}, Combustion and Flame 156 (2009), pp. 1035--1045,
  \url{https://doi.org/10.1016/j.combustflame.2008.10.029}.

\bibitem{AspdenJFM:2011}
A.J. Aspden, M.S. Day, and J.B. Bell, \emph{Turbulence-flame interactions in
  lean premixed hydrogen: {T}ransition to the distributed burning regime}, J.
  Fluid Mech. 680 (2011), pp. 287--320,
  \url{https://doi.org/10.1017/jfm.2011.164}.

\bibitem{Aspden:2017}
A. Aspden, \emph{A numerical study of diffusive effects in turbulent lean
  premixed hydrogen flames}, Proc. Combust. Inst. 36 (2017), pp. 1997--2004,
  \url{https://doi.org/10.1016/j.proci.2016.07.053}.

\bibitem{Schlup2017}
J. Schlup and G. Blanquart, \emph{Validation of a mixture-averaged thermal
  diffusion model for premixed lean hydrogen flames}, Combustion Theory and
  Modelling 22 (2018), pp. 264--290,
  \url{https://doi.org/10.1080/13647830.2017.1398350}.

\bibitem{Coffee:1981}
T. Coffee and J. Heimerl, \emph{Transport algorithms for premixed, laminar
  steady-state flames}, Combust. Flame 43 (1981), pp. 273--289,
  \url{https://doi.org/10.1016/0010-2180(81)90027-4}.

\bibitem{Ern:1998}
A. Ern and V. Giovangigli, \emph{Thermal diffusion effects in hydrogen-air and
  methane-air flames}, Combust. Theor. Model. 2 (1998), pp. 349--372,
  \url{https://doi.org/10.1088/1364-7830/2/4/001}.

\bibitem{Ern:1999}
A. Ern and V. Giovangigli, \emph{Impact of detailed multicomponent transport on
  planar and counterflow hydrogen/air and methane/air flames}, Combust. Sci.
  Technol. 149 (1999), pp. 157--181,
  \url{https://doi.org/10.1080/00102209908952104}.

\bibitem{Bongers:2003}
H. Bongers and L. De~Goey, \emph{The effect of simplified transport modeling on
  the burning velocity of laminar premixed flames}, Combust. Sci. Technol. 175
  (2003), pp. 1915--1928, \url{https://doi.org/10.1080/713713111}.

\bibitem{Yang2010}
F. Yang, C. Law, C. Sung, and H. Zhang, \emph{{A mechanistic study of Soret
  diffusion in hydrogen–air flames}}, Combust. Flame 157 (2010), pp.
  192--200, \url{https://doi.org/10.1016/j.combustflame.2009.09.018}.

\bibitem{Xin2012}
Y. Xin, C.J. Sung, and C.K. Law, \emph{{A mechanistic evaluation of Soret
  diffusion in heptane/air flames}}, Combust. Flame 159 (2012), pp. 2345--2351,
  \url{https://doi.org/10.1016/j.combustflame.2012.03.005}.

\bibitem{Giovangigli2015MulticomponentFlames}
V. Giovangigli, \emph{{Multicomponent transport in laminar flames}}, Proc.
  Combust. Inst. 35 (2015), pp. 625--637,
  \url{https://doi.org/10.1016/j.proci.2014.08.011}.

\bibitem{Dworkin2009TheFlames}
S. Dworkin, M. Smooke, and V. Giovangigli, \emph{{The impact of detailed
  multicomponent transport and thermal diffusion effects on soot formation in
  ethylene/air flames}}, Proc. Combust. Inst. 32 (2009), pp. 1165--1172,
  \url{https://doi.org/10.1016/j.proci.2008.05.061}.

\bibitem{Xin:2015}
Y. Xin, W. Liang, W. Liu, T. Lu, and C.K. Law, \emph{A reduced multicomponent
  diffusion model}, Combust. Flame 162 (2015), pp. 68--74,
  \url{https://doi.org/10.1016/j.combustflame.2014.07.019}.

\bibitem{Fillo2021}
A.J. Fillo, J. Schlup, G. Blanquart, and K.E. Niemeyer, \emph{Assessing the
  impact of multicomponent diffusion in direct numerical simulations of
  premixed, high-karlovitz, turbulent flames}, Combustion and Flame 223 (2021),
  pp. 216--229, \url{https://doi.org/10.1016/j.combustflame.2020.09.013}.

\bibitem{Desjardins2008}
O. Desjardins, G. Blanquart, G. Balarac, and H. Pitsch, \emph{{High order
  conservative finite difference scheme for variable density low Mach number
  turbulent flows}}, J. Comput. Phys. 227 (2008), pp. 7125--7159,
  \url{https://doi.org/10.1016/j.jcp.2008.03.027}.

\bibitem{Savard2015AChemistry}
B. Savard, Y. Xuan, B. Bobbitt, and G. Blanquart, \emph{A
  computationally-efficient, semi-implicit, iterative method for the
  time-integration of reacting flows with stiff chemistry}, J. Comput. Phys.
  295 (2015), pp. 740--769, \url{https://doi.org/10.1016/j.jcp.2015.04.018}.

\bibitem{Fillo2020_jcp}
A.J. Fillo, J. Schlup, G. Beardsell, G. Blanquart, and K.E. Niemeyer, \emph{A
  fast, low-memory, and stable algorithm for implementing multicomponent
  transport in direct numerical simulations}, Journal of Computational Physics
  406 (2020), p. 109185, \url{https://doi.org/10.1016/j.jcp.2019.109185}.

\bibitem{Curtiss1949TransportMixtures}
C.F. Curtiss and J.O. Hirschfelder, \emph{Transport properties of
  multicomponent gas mixtures}, J. Chem. Phys. 17 (1949), pp. 550--555,
  \url{https://doi.org/10.1063/1.1747319}.

\bibitem{Hirschfelder1954}
J.O. Hirschfelder, C.F. Curtiss, and R.B. Bird, \emph{{Molecular Theory of
  Gases and Liquids}}, Wiley, New York, 1954.

\bibitem{Grcar2009TheFlames}
J.F. Grcar, J.B. Bell, and M.S. Day, \emph{{The Soret effect in naturally
  propagating, premixed, lean, hydrogen–air flames}}, Proc. Combust. Inst. 32
  (2009), pp. 1173--1180, \url{https://doi.org/10.1016/j.proci.2008.06.075}.

\bibitem{KeeCRF}
R.J. Kee, M.E. Coltrin, P. Glarborg, and H. Zhu, \emph{Chemically Reacting
  Flow}, 2nd ed., John Wiley \& Sons, Inc, 2018.

\bibitem{Giovangigli1991}
V. Giovangigli, \emph{Convergent iterative methods for multicomponent
  diffusion}, {IMPACT} of Computing in Science and Engineering 3 (1991), pp.
  244--276, \url{https://doi.org/10.1016/0899-8248(91)90010-r}.

\bibitem{Kee1989Chemkin-II:Kinetics}
R. Kee, F. Rupley, and J. Miller, \emph{{Chemkin-II}: A {Fortran} chemical
  kinetics package for the analysis of gas-phase chemical kinetics}, Sandia
  National Laboratories Report SAND89-8009 (1989).
  \url{https://doi.org/10.2172/5681118}.

\bibitem{Dixon-Lewis1968FlameSystems}
G. Dixon-Lewis, \emph{Flame structure and flame reaction kinetics. {II}.
  transport phenomena in multicomponent systems}, Proc. Royal Soc. A 307
  (1968), \url{https://doi.org/10.1098/rspa.1968.0178}.

\bibitem{Burali2016AssessmentFlows}
N. Burali, S. Lapointe, B. Bobbitt, G. Blanquart, and Y. Xuan,
  \emph{{Assessment of the constant non-unity Lewis number assumption in
  chemically-reacting flows}}, Combust. Theor. Model. 20 (2016), pp. 632--657,
  \url{https://doi.org/10.1080/13647830.2016.1164344}.

\bibitem{Lapointe2016FuelFlames}
S. Lapointe and G. Blanquart, \emph{{Fuel and chemistry effects in high
  Karlovitz premixed turbulent flames}}, Combust. Flame 167 (2016), pp.
  294--307, \url{https://doi.org/10.1016/j.combustflame.2016.01.035}.

\bibitem{Hong2011AnMeasurements}
Z. Hong, D.F. Davidson, and R.K. Hanson, \emph{{An improved H2/O2 mechanism
  based on recent shock tube/laser absorption measurements}}, Combust. Flame
  158 (2011), pp. 633--644,
  \url{https://doi.org/10.1016/j.combustflame.2010.10.002}.

\bibitem{Lam2013AAbsorption}
K.Y. Lam, D.F. Davidson, and R.K. Hanson, \emph{A shock tube study of {H}$_2$ +
  {OH} $\to$ {H}$_2${O} + {H} using {OH} laser absorption}, Int. J. Chem.
  Kinet. 45 (2013), pp. 363--373, \url{https://doi.org/10.1002/kin.20771}.

\bibitem{Hong2013OnAbsorption}
Z. Hong, K.Y. Lam, R. Sur, S. Wang, D.F. Davidson, and R.K. Hanson, \emph{On
  the rate constants of {OH} + {HO}$_2$ and {HO}$_2$ + {HO}$_2$: A
  comprehensive study of {H}$_2${O}$_2$ thermal decomposition using
  multi-species laser absorption}, Proc. Combust. Inst. 34 (2013), pp.
  565--571, \url{https://doi.org/10.1016/j.proci.2012.06.108}.

\bibitem{Savard2015}
B. Savard and G. Blanquart, \emph{Broken reaction zone and differential
  diffusion effects in high {Karlovitz} \textit{n}-{C}$_7${H}$_{16}$ premixed
  turbulent flames}, Combust. Flame 162 (2015), pp. 2020--2033,
  \url{https://doi.org/10.1016/j.combustflame.2014.12.020}.

\bibitem{Rosales2005}
C. Rosales and C. Meneveau, \emph{Linear forcing in numerical simulations of
  isotropic turbulence: Physical space implementations and convergence
  properties}, Phys. Fluids 17 (2005), p. 095106,
  \url{https://doi.org/10.1063/1.2047568}.

\bibitem{Carroll2013}
P.L. Carroll and G. Blanquart, \emph{A proposed modification to {Lundgren}'s
  physical space velocity forcing method for isotropic turbulence}, Phys.
  Fluids 25 (2013), p. 105114, \url{https://doi.org/10.1063/1.4826315}.

\bibitem{Klein2017}
M. Klein, N. Chakraborty, and S. Ketterl, \emph{A comparison of strategies for
  direct numerical simulation of turbulence chemistry interaction in generic
  planar turbulent premixed flames}, Flow, Turbulence and Combustion 99 (2017),
  pp. 955--971, \url{https://doi.org/10.1007/s10494-017-9843-9}.

\bibitem{Williams:1985}
F.A. Williams, \emph{Combustion Theory}, Benjamin/Cummings, 1985.

\bibitem{Aspden:2015}
A. Aspden, M. Day, and J. Bell, \emph{Turbulence-chemistry interaction in lean
  premixed hydrogen combustion}, Proc. Combust. Inst. 35 (2015), pp.
  1321--1329, \url{https://doi.org/10.1016/j.proci.2014.08.012}.

\bibitem{Steinberg2021}
A.M. Steinberg, P.E. Hamlington, and X. Zhao, \emph{Structure and dynamics of
  highly turbulent premixed combustion}, Progress in Energy and Combustion
  Science 85 (2021), p. 100900,
  \url{https://doi.org/10.1016/j.pecs.2020.100900}.

\bibitem{Geikie2018}
M.K. Geikie and K.A. Ahmed, \emph{Pressure-gradient tailoring effects on the
  turbulent flame-vortex dynamics of bluff-body premixed flames}, Combustion
  and Flame 197 (2018), pp. 227--242,
  \url{https://doi.org/10.1016/j.combustflame.2018.08.001}.

\bibitem{Kazbekov2019}
A. Kazbekov, K. Kumashiro, and A.M. Steinberg, \emph{Enstrophy transport in
  swirl combustion}, Journal of Fluid Mechanics 876 (2019), pp. 715--732,
  \url{https://doi.org/10.1017/jfm.2019.551}.

\bibitem{Kazbekov2021}
A. Kazbekov and A.M. Steinberg, \emph{Flame- and flow-conditioned vorticity
  transport in premixed swirl combustion}, Proceedings of the Combustion
  Institute 38 (2021), pp. 2949--2956,
  \url{https://doi.org/10.1016/j.proci.2020.06.211}.

\bibitem{Kim2007}
S.H. Kim and H. Pitsch, \emph{Scalar gradient and small-scale structure in
  turbulent premixed combustion}, Phys. Fluids 19 (2007), p. 115104,
  \url{https://doi.org/10.1063/1.2784943}.

\bibitem{repropack}
A.J. Fillo, P.E. Hamlington, and K.E. Niemeyer, \emph{Figures, plotting
  scripts, and data for ``{Assessing} diffusion model impacts on turbulent
  transport and flame structure in lean premixed flames'' [dataset]}, Zenodo
  (2022). \url{https://doi.org/10.5281/zenodo.6191166}.

\bibitem{vorticityData}
A.J. Fillo, P.E. Hamlington, and K.E. Niemeyer, \emph{Assessing the impact of
  diffusion model on the turbulent transport and flame structure of premixed
  lean hydrogen flames: hydrogen data (version 1) [dataset]}, Oregon State
  University (2020). \url{https://doi.org/10.7267/37720k356}.

\end{thebibliography}

\appendix
\section{Availability of material}

The figures in this article, as well as the data and plotting scripts necessary 
to reproduce them, are available openly under the CC-BY license~\cite{repropack}.
Furthermore, the full simulation inputs for and output data produced by NGA are 
available~\cite{vorticityData}.

\end{document}